\documentclass[11pt,b5paper]{article}
\usepackage{amsmath}
\usepackage{amsthm}
\usepackage{amssymb}
\usepackage{latexsym}
\usepackage{amsfonts}
\newcommand{\nc}{\newcommand*}

\nc{\ts}{\textstyle}
\nc{\ds}{\displaystyle}
\nc{\wt}{\widetilde}
\def\half{{\frac{1}{2}}}
\def\BI{{\Bbb I}}
\def\ta{\wt{a}}
\def\tF{\wt{F}}
\def\tH{\wt{H}}
\def\tK{\wt{K}}
\def\tP{\wt{P}}
\def\tX{\wt{X}}
\def\cF{{\mathcal F}}
\def\cH{{\mathcal H}}
\def\cN{{\mathcal N}}
\def\Kn{K_n(x;p,N)}
\def\tKn{{\tK}_n(x;p,N)}
\def\tKzn{{\tK}_n^{(0)}(x;p,N)}
\def\HS{\widetilde{\cH}_{p,N}}
\nc{\lan}{\langle}
\nc{\ran}{\rangle}
\nc{\bra}[1]{\lan{#1}\vert}                
\nc{\ket}[1]{\vert{#1}\ran}                
\nc{\braket}[2]{{\langle{#1}\vert{#2}\rangle}}  
\nc{\tKx}[1]{{\tK}_{#1}(x;p,N)}
\nc{\tKzx}[1]{{\tK}_{#1}^{(0)}(x;p,N)}
\nc{\tghfto}[3]{{{}_{2}{\tF}_{1}}\left(\ts{\genfrac{}{}{0pt}{}{#1}{#2}\biggl.\biggr| #3 }\right)}
\nc{\Szx}[2]{\sum_{ #1 = 0 }^{#2}}
\nc{\reff}[1]{(\ref{#1})}
\begin{document}

\title{Coherent states of Krawtchouk oscillator and beyond}
{V.V.Borzov}
{Department of Mathematics, St.Petersburg University of
Telecommunications,
191065, Moika  61, St.Petersburg, Russia}
{vadim@VB6384.spb.edu}
and
{E. V. Damaskinsky}
{Department of Mathematics, University of Defense
Technical Engineering,
191123, Zacharievskaya 22, St.Petersburg, Russia}
{evd@pdmi.ras.ru}

\vspace{1cm}
\centerline{\bf Abstract}
\begin{quote}
In the frame of our approach we constructed the generalized oscillator
connected with Krawtchouk polynomials (named Krawtchouk oscillator) and
coherent states for this oscillator too. Ours results are compared with
analogues ones obtained for another variant of Krawtchouk oscillator
in the paper [N.M.Atakishiev and S.K.Suslov, {\it Difference analogs 
of the harmonic oscillator}, Teor. Mat. Fiz., {\bf 85}, 64-73 (1990)] 
from other point of view. Our definition of coherent
states is close to one given in [B.Roy and P.Roy, {\it Phase properties 
of a new nonlinear coherent state}, quant-ph/0002043]\footnote{This 
investigation is partially supported by RFBR grant No 06-01-00451}
\bigskip
\end{quote}

{\bf 1. Introduction.} The Krawtchouk polynomials (\cite{1} - \cite{3})
have interesting applications in the quantum optics (for instance see the
oscillator model studied in the papers \cite{4}, \cite{5}). These
polynomials provide an important example of the classical orthogonal
polynomials of discrete variable. It is very attractive for some
applications that they can be considered as finite-dimensional
approximations of the Hermite and Charlier polynomials \cite{3}.
In the papers \cite{6} - \cite{11} the authors proposed a new approach
to construction of oscillator-like systems (so-called generalized
oscillators) connected with given family of orthogonal polynomials.
Furthermore, in these papers was introduced and investigated the
coherent states for this type oscillators. Now in the frame of our
approach we construct the generalized oscillator connected with
Krawtchouk polynomials (named Krawtchouk oscillator) and coherent states
for this oscillator too. Our results are consistent with analogous ones
obtained in the papers \cite{4}, \cite{5}, in which similar oscillator
was considered from other point of view.

The most attention in our paper is given to construction of
coherent states. Because of the space of the states for Krawtchouk
oscillator is finite-dimensional space it is obvious that the
annihilation operator
cannot have any eigenvalue differ from zero. This means that in this
case the standard definition of coherent states as eigenvectors of
this operator (the coherent states of Barut - Girardello type)
falls. In this connection we generalize
the method of the construction of coherent states that goes back to the
Glauber's papers \cite{Gl}. This approach was used in the case of so-called
finite-dimensional (truncated) variant of the standard boson oscillator
in the paper \cite{12}. We compare the resulting coherent states for
Krawtchouk oscillator with ones considered in the papers
\cite{4}, \cite{13}, \cite{14}.

\bigskip\bigskip

\noindent{\bf 2. Krawtchouk Polynomials.} Let us recall the definition
of Krawtchouk polynomials (\cite{1} - \cite{3})
\begin{equation}
\Kn:=\tghfto{-n,-x}{-N}{p^{-1}}=\Szx{k}{N}\frac{(-n)_k(-x)_k}{k!(-N)_k}p^{-k}
\end{equation}
where $0<p<1,$\quad $n=0,1,\ldots,N,$ and
\begin{equation}
(a)_0=1,\quad (a)_k=a(a+1)\cdot\ldots\cdot(a+k-1)=
\frac{\Gamma(a+k)}{\Gamma(a)} .
\end{equation}
It is convenient for construction of generalized oscillator to
renormalize the Krawtchouk polynomials, following \cite{6}
\begin{equation}
\tKn=\sqrt{\rho(n,p,N)}\Kn ,\quad n=0,1,\ldots,N ,
\end{equation}
where
\begin{equation}
\rho(n,p,N)=C_N^{n}p^{n}(1-p)^{N-n} , \qquad
C_N^{n}:=\frac{N!}{\Gamma(n+1)\Gamma(N-n+1)} .
\end{equation}
The renormalized Krawtchouk polynomials $\wt{K}_n$ fulfill recurrent
relations with symmetrical Jacobi matrix
\begin{align}
\!\!\!x\tKn\!=\!b_n\tKx{n+1}&\!+\!a_n\tKn\!+\!b_n\tKx{n-1},
\!\quad n\!=\!0,1,\ldots,d ,\label{5}
\\
\tKx{0}&=1 ,
\end{align}
where
\begin{equation}\label{7}
a_n=p(N-n)+n(1-p),\qquad b_n=-\sqrt{p(1-p)(n+1)(N-n)} ,
\quad n=0,1,\ldots,N.
\end{equation}
The Krawtchouk polynomial $\tKn$ is a solution of the difference equation
\begin{multline}
-n\tKn(x;p,N)=p(N-x)\tKn(x+1;p,N)-\\
-\left[p(N-x)+x(1-p)\right]\tKn(x;p,N)+\\
+x(1-p)\tKn(x-1;p,N)\label{8}
\end{multline}
We recall also orthogonality relations for Krawtchouk polynomials
\begin{align}
\Szx{x}{N}\rho(x;p,N)\tKx{m}\tKn=\delta_{m,n};\\
\Szx{x}{N}\rho(n;p,N)\tKn{\tK}_n(y;p,N)=\delta_{x,y} ,
\end{align}
where $0<p<1$ and $n,x=0,1,\ldots,N.$

{\bf 3. Krawtchouk Oscillator.}
Now we describe (following \cite{6}) the construction of Krawtchouk
oscillator. Let $\HS=\ell^2_{N+1}(\rho(x;p,N))$ be the $N+1$-dimensional
Hilbert space spanned by the orthonormal basis
$\left\{\tKn\right\}_{n=0}^N$ (with weight function $\rho(x;p,N)$).
We call the Krawtchouk oscillator the oscillator - like system
in $\HS$ defined by generalized operators of "coordinate" $\tX,$
"momentum" $\tP$ and quadratic Hamiltonian $\tH$
\begin{align}
\tX:&=\text{Re}(X-P),\\
\tP:&=-i\text{Im}(X-P),\\
\tH:&=\frac{1}{4p(1-p)}\left(\tX^{\,2}+\tP^{\,2}\right).
\end{align}
Here the selfadjoint operators $X$ and $P$ are defined by
the action on basis elements $\tKx{n}$ in the Hilbert space $\HS$
\begin{align}
X\tKx{n}&\!=\!b_n\tKx{n+1}\!+\!a_n\tKx{n}\!+\!b_{n-1}\tKx{n-1},\\
P\tKx{n}&\!=\!-ib_n\tKx{n+1}\!+\!a_n\tKx{n}\!+\!ib_{n-1}\tKx{n-1},
\end{align}
\vspace{-0.6cm}
\begin{align}
X\tKx{0}&=b_0\tKx{1}+a_0\tKx{0},\\
P\tKx{0}&=-ib_0\tKx{1}+a_0\tKx{0},
\end{align}
where
\begin{equation}
a_n=p(N-n)+n(1-p),\qquad b_n=-\sqrt{p(1-p)(n+1)(N-n)} ,
\quad n=0,1,\ldots,N ,
\end{equation}

The action of creation and annihilation operators
\begin{equation}
\ta^{\pm}:=\frac{1}{2\sqrt{p(1-p)}}\left(\tX\pm i\tP\right),
\end{equation}
on the basis states in $\HS$ are given by the relations
\begin{align}
\ta^{-}\tKx{n}&=-\sqrt{n(N-n+1)}\tKx{n-1},\label{21}\\[8pt]
\ta^{+}\tKx{n}&=-\sqrt{(n+1)(N-n)}\tKx{n+1}.\label{22}
\end{align}
These operators fulfill commutation relations
\begin{equation}
\left[\ta^{-}, \ta^{+}\right]=(N-1)\BI-2\cN
\end{equation}
where operator $\cN$ is defined by
\begin{equation}
\cN\tKx{n}=n\tKx{n} .
\end{equation}

From the relations \reff{21}-\reff{22} it follows that
the eigenvalues of Hamiltonian operator
\begin{equation}
\tH=\half\left(\ta^{+}\ta^{-}+\ta^{-}\ta^{+}\right)
\end{equation}
are given by  $ (0\leq n\leq N)$
\begin{equation}
\tH\tKx{n}=\lambda_n\tKx{n},\quad
\lambda_n=N(n+{\ts\half})-n^2,
\end{equation}
so that $\lambda_N=\lambda_0={\half}N.$ Using the results obtained
in \cite{7} it can be shown that difference equation \reff{8}
for Krawtchouk polynomials $\tKx{n}$ is equivalent to eigenvalue
equation for Hamiltonian $\tH$ in the space $\HS.$ For obtaining
the explicit expression of this operator as a difference operator
in $\HS$ it is convenient to compare our variant of Krawtchouk
oscillator with one was considered in \cite{4}.

\bigskip
{\bf 4. Comparison of Krawtchouk oscillator with variant considered in
the work of Atakishiev and Suslov \cite{4}.}
In the works \cite{4}, \cite{5} Atakishiev and Suslov studied
the Krawtchouk oscillator with Hamiltonian
\begin{equation}\label{27}
H_{AS}=2p(1-p)N+\half+(1-2p)\frac{\xi}{h}-\sqrt{p(1-p)}
\left[\alpha(\xi)e^{h\partial_{\xi}}+
\alpha(\xi-h)e^{-h\partial_{\xi}}\right],
\end{equation}
where
\begin{equation}\label{28}
h=\sqrt{2Np(1-p)},\qquad\alpha(\xi)=
\sqrt{\left((1-p)N-\frac{\xi}{h}\right)\left(pN+1+\frac{\xi}{h}\right)}.
\end{equation}
This operator is defined in the Hilbert space $\cH_{AS}=\ell^2(\xi)$
with orthonormal basis consisting of Krawtchouk functions
\begin{equation}\label{29}
\Psi_n(\xi)=(-1)^n\sqrt{C_N^n\left(\frac{p}{1-p}\right)^n
\rho(pN+\frac{\xi}{h};p,N)}K_n(pN+\frac{\xi}{h};p,N),
\end{equation}
which fulfill the orthogonality relations ($\xi_j=h(j-pN)$)
\begin{equation}\label{30}
\sum_{j=0}^{N}\Psi_n(\xi_j)\Psi_m(\xi_j)=\delta_{n,m},\qquad
\sum_{n=0}^{N}\Psi_n(\xi_i)\Psi_n(\xi_j)=\delta_{i,j}.
\end{equation}
The Krawtchouk functions are eigenfunctions of Hamiltonian $H_{AS}$
in the space $\cH_{AS}$
\begin{equation}\label{31}
H_{AS}\Psi_n(\xi)=\lambda_n\Psi_n(\xi),\qquad \lambda_n=n=\half,
\quad n=0,1,\ldots,N.
\end{equation}

The Hamiltonian $H_{AS}$ can be factorized
\begin{equation}\label{32}
H_{AS}(\xi)=\half\left[A^+,A^-\right]+\half(N+1),
\end{equation}
by ladder operators
\begin{align}
A^+(\xi)=(1-p)e^{-h\partial_{\xi}}\alpha(\xi)-p\alpha(\xi)e^{h\partial_{\xi}}
+\sqrt{p(1-p)}\left((2p-1)N+\frac{2\xi}{h}\right),\label{33} \\
A^-(\xi)=(1-p)\alpha(\xi)e^{h\partial_{\xi}}-pe^{-h\partial_{\xi}}\alpha(\xi)
+\sqrt{p(1-p)}\left((2p-1)N+\frac{2\xi}{h}\right),\label{34}
\end{align}
which act on the basis elements according to
\begin{equation}\label{35}
A^+(\xi)\Psi_n(\xi)=\sqrt{(n+1)(N-n)}\Psi_{n+1}(\xi),\qquad
A^-(\xi)\Psi_n(\xi)=\sqrt{n(N-n+1)}\Psi_{n-1}(\xi).
\end{equation}

The operators $A^{\pm}$ and the operator
\begin{equation}\label{36}
A_0(\xi):=\half\left[A^+(\xi),A^-(\xi)\right]
\end{equation}
fulfill the commutation relations of the Lee algebra $so(3)$
\begin{equation}\label{37}
\left[A^+(\xi),A^-(\xi)\right]=2A_0(\xi),\qquad
\left[A_0(\xi),A^{\pm}(\xi)\right]={\pm}A^{\pm}(\xi).
\end{equation}

Let us denote by $\tH_{AS}$ the selfadjoint operator in the Hilbert space
$\cH_{p,N}$ that unitary equivalent to the selfadjoint operator $H_{AS}$
in the space $\cH_{AS}$
\begin{equation}\label{38}
\tH_{AS}=T^{-1}H_{AS}T.
\end{equation}
The  explicit expression for unitary operator
$T:\cH_{p,N}\rightarrow\cH_{AS}$
is rather cumbersome. For brevity we omit it.
It follows from comparing the spectrum of the
hamiltonian operators $\tH$ and $\tH_{AS}$ that these operators
are connected by the relation
\begin{equation}\label{39}
\tH=-\left(\tH_{AS}-\half\BI\right)^2+N\tH_{AS}.
\end{equation}

\bigskip
{\bf 5. Coherent states for Krawtchouk oscillator.}
We consider the $(N+1)$-dimensional Hilbert space $\cH_{p,N}$ as
the Fock space $\cF$ equipped with the basis states
$\ket{0},\ket{1},\ldots,\ket{N}.$ Note that in coordinate picture
we have $\braket{x}{n}=\tKx{n}.$

We define the coherent states (of the Glauber type) in $\cF$
by the relation
\begin{equation}\label{40}
\ket{z}=\Szx{n}{N}\frac{\left(z\ta_+-\bar{z}\ta^-\right)^n}{n!}\ket{0}.
\end{equation}
One can rewrite this definition in the following form
(wich is close to the coherent state considered in \cite{12})
\begin{equation}\label{41}
\ket{z}=\Szx{l}{N}\sum_{n=l}^{\infty}d_{n,l}^{N}
\frac{\left(\sqrt{2}b_{l-1}\right)!}{n!}(-\bar{z})^{\half(n-l)}
(z)^{\half(n+l)}\ket{l}
\end{equation}
where $b_k$ are given by \reff{7} and coefficients  $d_{n,l}^{N}$
fulfill the recurrence relations
\begin{align}
d_{n,l}^{N}&=\theta_{l}d_{n-1,l-1}^{N}
+2b_{l}^{\,\,2}\theta_{l+1}d_{n-1,l+1}^{N-1},\\
d_{n-1,-1}^{N}&=0,\qquad d_{0,0}^{N}=1,\qquad d_{n,n+k}^{N}=0
\quad\text{for}\quad k>0.
\end{align}
The solution of these recurrence relations has the form
\begin{equation}\label{44}
d_{n,l}^{N}=
\frac{1}{\left(2b_{l-1}^{\,\,2}\right)!\Szx{k}{N}
\left(\wt{\psi}_N^{\,-2}(x_k)\right)}
\Szx{k}{N}\frac{\wt{\psi}_l(x_k)}{\left(\wt{\psi}_N(x_k)\right)^2},
\end{equation}
here $x_k$ ($k=0,1,\ldots,N)$ are the roots of the equation
\begin{equation}\label{45}
\wt{\psi}_{N+1}(x)=0,
\end{equation}
where
\begin{equation}\label{46}
\wt{\psi}_{n}(x\sqrt{2})=\left(\sqrt{2}b_{n-1}\right)!\tKzn,
\end{equation}
and polynomials $\tKzn$ fulfill the recurrent relations
\begin{align}
\!\!\!x\tKzn\!=\!b_n\tKzx{n+1}&\!+\!b_{n-1}\tKzx{n-1},
\quad n=0,1,\ldots,N ,\label{47}
\\
\tKzx{0}&=1 ,\label{48}
\end{align}
which can be obtained from \reff{5} if we put $a_n=0.$

With help of the relations \reff{44} and \reff{46} we can rewrite
the expression for the coherent states in the form
\begin{equation}
\ket{z}=
\frac{1}{\Szx{k}{N}\left(\wt{\psi}_N^{-2}(x_k)\right)}
\Szx{l}{N}\left(-\frac{iz}{|z|}\right)^l
\frac{1}{\left(\sqrt{2}b_{l-1}\right)!}
\left(\sum_{k=0}^{N}\frac{{\tK}_{l}(x_k;p,N)}{({\tK}_{N}(x_k;p,N))^2}
e^{i|z|x_k}\right)\ket{l} \label{49}
\end{equation}

\bigskip
{\bf 5. Comparison with other types of coherent states.}
Let us compare definition of coherent states given above with some
other definitions of coherent states for Krawtchouk oscillator:
the so-called spin coherent states \cite{4}, \cite{13} and
phase - type coherent states \cite{14}.

The spin coherent states for Krawtchouk oscillator
in papers \cite{4}, \cite{13} was given in the form
\begin{equation}\label{51}
\braket{x}{\xi}=\left(1+|\xi|^2\right)^{-N/2}
\sum_{n=0}^{N}\sqrt{C_N^n}\xi^n\Psi_n(x),
\end{equation}
where $\Psi_n(x)$ defined by \reff{29}.
It is obvious that these states differ from the states \reff{49}.

The phase coherent states suggested in \cite{14} look as
\begin{equation}\label{52}
\ket{z}=\left(1+\left|\frac{2\pi z}{N+1}\right|^2\right)^{-N/2}
\sum_{n=0}^{N}\sqrt{C_N^n}\left(\frac{2\pi z}{N+1}\right)^n
\ket{\theta_n},
\end{equation}
where
\begin{equation}\label{53}
\ket{\theta_n}=(N+1)^{-\half}\sum_{k=0}^{N}e^{ik\theta_n}\ket{k},\qquad
\theta_n=\theta_0+\frac{2\pi n}{N+1},\quad n=0,1,\ldots,N+1.
\end{equation}
These states are also different from the states \reff{49} given above.

Our coherent states \reff{49} similar to the so called
finite-dimensional coherent states suggested in \cite{15},\cite{12}
in the framework of the Pegg - Barnet formalism. We plain to
consider the main properties of coherent states \reff{49}  in
other publication.

\end{document}